\documentstyle [12pt] {article}
\begin{document}
\begin{titlepage}
\begin{center}
\vspace{2cm}
\LARGE
The Ages of Elliptical Galaxies in a Merger Model \\
\vspace{1cm}
\large
Guinevere Kauffmann \\
\vspace{0.5cm}
{\em Max Planck Institut fur Extraterrestrische Physik, D 85740 Garching bei
Muenchen,
Germany}\\
\vspace{0.8cm}
\end{center}
\large
\begin {abstract}
The tightness of the observed colour-magnitude and Mg$_{2}$- velocity
dispersion relations for
elliptical galaxies has often been cited as an argument against a picture in
which ellipticals form
by the merging of spiral disks. A common view is that merging would mix
together stars of
disparate ages and produce a large scatter in these relations.
Here I use semi-analytic models of galaxy formation to derive the distribution
of the mean ages,
colours and metallicities of the stars in elliptical galaxies formed by mergers
in a flat CDM
universe. It is seen that most of the stars in ellipticals form at relatively
high redshift
($z>1.9$) and that the predicted scatter in the colour-magnitude and
Mg$_{2}$-$\sigma$
relations falls within observational bounds. I conclude that the apparent
``homogeneity''
in the properties of the stellar populations of ellipticals is not inconsistent
with a merger
scenario for the origin of these systems.\\

\vspace {2cm}
{\em Subject Headings:} Galaxies: Elliptical -- Galaxies: Formation --
Galaxies: Stellar Content --
 Galaxies: Fundamental Parameters
\end {abstract}
\end {titlepage}

\section {Introduction}
There have long been two competing views on the formation history of the
ellipticals galaxies
we see today. One is that most of the stars in present-day galactic bulges and
ellipticals
were produced during a relatively short, early phase of intense star formation
at high redshift.
The second view is that elliptical galaxies are relative latecomers, having
been produced as the
result of the merging of disk galaxies drawn together by gravity
as their surrounding dark matter halos coalesced.

In recent years, the second view has come to enjoy increasing popularity.
Detailed numerical
simulations have shown that mergers between two spiral galaxies of comparable
mass lead to
the production of spheroidal merger remnants with physical characteristics such
density profiles, gravitational radii, mean velocity dispersions and surface
brightnesses that are
quite comparable to observed ellipticals (see Barnes \& Henquist 1992 for a
review).
In addition, observational eveidence has begun to
accumulate that mergers and interactions are rather common in the universe,
particularly
at higher redshift. Hubble Space Telescope images indicate that at redshifts
$\simeq 0.4$,
25-30 percent of all galaxies show disturbed morphologies (Griffiths {\em et
al} 1995).
At lower redshifts, almost all the
most luminous IRAS galaxies have turned out to be interacting systems, leading
to the conclusion
that the merging process may fuel very intense bursts of star formation (see
Joseph 1990 for a
review). Schweizer \& Seizer (1992) present evidence indicating that the UBV
colours of elliptical
galaxies become systematically bluer as the amount of fine structure in these
galaxies increases.
If fine structure is indeed a measure of a galaxy's dynamical youth, these
observations
suggest that many of today's ellipticals have undergone a merger-induced
starburst in the past.
Likewise, analyses of the stellar ages of elliptical galaxies using
absorption-line
index strengths, show that the outer parts of elliptical galaxies are
both older and more metal poor than the nuclei, consistent with the
merger/starburst picture.

The fact remains, however, that the stellar populations of present-day
elliptical galaxies
appear to be rather homogeneous. The colours, mass-to-light ratios and Mg$_{2}$
absorption
strength of ellipticals are correlated only with their central velocity
dispersions, and the
scatter about these relations appears to be surprisingly small. Bower, Lucey \&
Ellis (1992)
show that the intrinsic scatter about the U-V and V-K colour-magnitude
relations for early-type
galaxies in the Coma and Virgo clusters is less than 0.05 mag. This leads to
the conclusion
that elliptical galaxies must have formed the bulk of their stars at redshifts
greater than 2, or
else the observed homogeneity in the colours would require a very precise
synchronization
of formation epochs and star formation histories for these galaxies. Likewise,
Bender, Burstein
\& Faber (1993) find a very tight relationship between the strength of the
Mg$_{2}$ index at
the centre of ellipticals and their central velocity dispersions, $\sigma$.
Galaxies show a mean
scatter of only 0.025 mag in Mg$_{2}$, versus a full range of nearly 0.35 mag
in Mg$_{2}$.
This observed scatter sets an upper limit of 15\% on the rms variation of both
age and
metallicity at fixed $\sigma$ for bright ellipticals.

The tightness of these limits calls into question the tenability of the merger
model, as merging might
be expected to mix together stellar populations of quite disparate ages
(Renzini 1994).
In addition, a starburst accompanying the merging event would increase the
amount of light
contributed by young stars. However, detailed predictions of the resulting
scatter in quantities such
as  colours and Mg$_{2}$ strengths have not yet been made for any realistic
theory of
elliptical galaxy formation via merging.
In this letter, I present results calculated
for a $b=1.5$ cold dark matter (CDM) universe using semi-analytic techniques.
It has already
been demonstrated
in Kauffmann (1995) that this model provides a good fit to both the properties
of
cluster ellipticals seen today, and the evolution in the colours of cluster
galaxies at
higher redshift. The results of the analysis here show that although elliptical
formation
is far from a synchronized process in this model,
merging does in fact take place at early enough epochs
so that the scatter in the colours and mean stellar ages are in good agreement
with
observations.

\section {Semianalytic Models of Elliptical Galaxy Formation}
The semi-analytic models we employ are described in detail in Kauffmann, White
\& Guiderdoni (1993, hereafter KWG) and Kauffmann \& White (1993). Application
of the model to the
evolution of the galaxy population in clusters at high redshift is discussed in
Kauffmann (1995).

To summarize:\\
I use an algorithm based on an extension of the Press-Schechter theory due to
Bower (1991) and
Bond {\em et al} (1991) to generate Monte
Carlo realizations of the merging paths of dark matter halos from high
redshift until the present. Dark matter halos are modelled as truncated
isothermal spheres and
it is assumed that as the halo forms, the gas relaxes to a distribution that
exactly
parallels that of the dark matter.
Gas then cools and condenses onto a {\em central galaxy} at the core of each
halo. Star formation and feedback processes take place as described in KWG.
In practice, star formation rate in central galaxies takes place at a roughly
constant rate of a few
solar masses per year, in agreement with the rates derived by Kennicutt (1983)
for normal spiral
galaxies.

At a subsequent redshift, a halo will
have merged with a number of others, forming a new halo of larger
mass. All gas which has not already cooled is assumed to be shock
heated to the virial temperature of this new halo. This hot gas then
cools onto the central galaxy of the new halo, which is identified
with the central galaxy of its {\em largest progenitor}. The central
galaxies of the other progenitors become {\em satellite galaxies},
which are able to merge with the central galaxy on a dynamical friction
timescale. If  a merger takes place between two galaxies of roughly comparable
mass, the merger remnant is labelled as an ``elliptical'' and all cold gas
is tranformed instantaneously into stars in a ``starburst''.

Note that the infall of new gas onto
satellite galaxies is not allowed, and star
formation will continue in such objects only until their existing cold gas
reservoirs are exhausted. Thus the epoch at which a galaxy is accreted by a
larger
halo delineates the transition between active star formation in the galaxy
and passive evolution of its stellar population. The stellar populations of
elliptical merger remnants in clusters hence redden as their stellar
populations age.
Central galaxy merger remnants in the ``field'' are able to accrete new gas in
the form of a disk
to form a ``spiral'' galaxy consisting of both a spheroidal bulge and a disk
component. As demonstrated in KWG, this picture is able to account for the
observed numbers and luminosity distributions of galaxies of different
morphologies, both in clusters and
in lower-density environments.

The spectrophotometric models of Bruzual \& Charlot (1993) are used to tranlate
the
predictions of the models into observed quantities such as magnitude and
colour, which
may be compared directly with the observational data.
In this letter, I focus on the ages and colours of the stars of the
elliptical merger remnants as predicted by the model.
The cosmological initial conditions are a $b=1.5$ CDM universe with $\Omega=1$
and
$H_{0}= 50$ km s$^{-1}$ Mpc$^{-1}$.

\section {The Ages and Colours of Elliptical Galaxies}

Figure 1 shows the V luminosity-weighted mean stellar age of the galaxies in a
cluster of $10^{15} M_{\odot}$. The symbols in the plot represent the
morphological
type of the galaxy: large filled circles correspond to ellipticals, open
squares are
SOs, small filled circles are spirals and pinwheels denote ellipticals which
have
been formed by a recent merging event ($<$ 1 Gyr ago). The classification into
morphological type is based on the B-band disk-to-bulge ratio of the galaxy, as
given in Table 3B of Simien \& de Vaucouleurs (1986).
As can be seen, the stellar populations of spiral galaxies span a very wide
range in
age, but those of early-type galaxies are confined to ages between 8 and 12.5
Gyr.
For our adopted cosmological parameters, this means that the bulk of stars in
elliptical and SO galaxies were formed at redshifts exceeding 1.9.

The age distribution of early-type galaxies found in halos in the mass range
$10^{12}-10^{13} M_{\odot}$ is plotted in figure 2a. These galaxies either
occur in
small groups, or have no companion of comparable luminosity, and thus can be
regarded as
``field'' ellipticals. In order to get a sizeable sample of these objects, it
was necessary
to combine the populations of 30 such low-mass halos. The faint field
ellipticals
have ages comparable to early-type galaxies in clusters. At brighter
magnitudes,
early-type galaxies tend to have considerably younger stellar populations. This
is because these objects
are all central galaxies, which continuously accrete gas from the surrounding
halo and thus manufacture stars right up to the present day.
Figure 2b shows results for galaxies in more massive groups ($10^{14}
M_{\odot})$.

Table 1 lists the mean stellar ages and the rms standard deviation for the
ellipticals and the ellipticals+SOs in both clusters and groups.
Ellipticals in all environments have mean ages between 10.5-11 Gyr with a
scatter
in age of around 10 percent. If all early-type galaxies are considered, the
scatter
increases to 15 percent and the mean stellar age is slightly lower.

Figure 3 is a histogram of the fraction of the integrated V-band light of a
cluster elliptical or
spiral galaxy contributed by stars which form at a given epoch. The solid line
represents the average
contribution, while
the filled squares and cicles represent the
maximum and minimum contributions respectively.
The difference between the average
star formation history of a cluster elliptical and spiral is illustrated
clearly in this
diagram. A large fraction of the V-band light of spirals comes from stars that
formed in
the past 5 Gyr, whereas {\em on average}, this contribution is negligible in
ellipticals.
It should be noted, however, that a small subset of elliptical galaxies do
exhibit
substantial star formation at late epochs. The dotted line in the top panel of
figure 3 illustrates the
epoch at which elliptical galaxies typical underwent their last major merging
event.
As can be seen, most ellipticals are formed by a merger that occurred between 5
and 10 Gyr ago.

Finally, in figure 4, I have used the evolutionary synthesis models of Bruzual
\& Charlot
(1993) to generate a scatterplot of the U-V and V-K colours of early-type
galaxies in clusters.
In U-V, the average colour of an
elliptical is 1.37 with a rms standard deviation, $\sigma$, of 0.034. In V-K,
the mean is
3.01 and $\sigma$ is 0.045. The rms is calculated using the Median Absolute
Difference
technique so that the results may be compared directly with the data of Bower,
Lucey \& Ellis (1992).
If SO galaxies are included, the rms increases
to $\simeq$ 0.07.
Although the mean colours are roughly comparable to the average colours in the
data of
Bower {\em et al}, there is no correlation bewteen the colour and the magnitude
of the galaxy, in contradiction with the observations, which show strong
reddening in the
stellar populations of elliptical galaxies with increasing luminosity.
As shown in figure 1, cluster elliptical galaxies of all
luminosities have roughly the same mean stellar age, so any trend in colour
would have to
result from metallicity differences between galaxies of different masses.
The rms scatter in the colour of the model ellipticals is well within  the
upper limit for the intrinsic
scatter in the colour of Coma and Virgo ellipticals quoted by Bower, Lucy \&
Ellis (1992).
The inclusion of the SO population increases the scatter
by a factor of two. It is interesting that
Bower {\em et al} also found that the removal of SO galaxies caused the
scatter to decrease in their data, but not by as much as a factor of 2.
However, it is clear that in the model, the scatter quoted for the early-type
population will depend on the precise value of the disk-to-bulge ratio at which
a galaxy is
labelled as being ``early-type''.

\section {A Simple Chemical Enrichment Model}
Faber (1973) discovered that metal line strengths (eg Mg$_{2}$) increase with
luminosity among
elliptical galaxies and interpreted this rise a due to an increase in
metallicity with increasing
mass. All subsequent studies that have analyzed Mg$_{2}$ using good data and
large samples, have
shown that the relationship between Mg$_{2}$ and the central velocity
dispersion $\sigma$ is
tight for bright ellipticals (Terlevich {\em et al} 1981; Dressler 1984;
Dressler {\em et al} 1987;
Burstein {\em et al} 1987; Bender, Burstein \& Faber 1993 (BBF)). The residuals
of Mg$_{2}$ on $\sigma$ have
a Gaussian core with a standard deviation of around 0.02 mag. Fainter galaxies
tend to have higher dispersion than the brighter ones.

This scatter can be translated into age or metallicity variations using
population synthesis models.
BBF derive the following approximate relation:
\begin {equation} \log Mg_{2} = 0.41 \log (Z/Z_{\odot}) + 0.41 \log (t/Gyr)
-1.00 \end {equation}
where $Z$ denotes metallicity and $t$ denotes age. Thus if the intrinsic
dispersion in Mg$_{2}$
is attributed solely to an age dispersion, the rms spread in age at fixed
$\sigma$ must be less
than 15 percent. Likewise, if the Mg$_{2}$ dispersion is due to metallicity
alone, this implies
that the scatter in $Z$ is also about 15\%. In the previous section, it was
demonstrated that the
model prediction for the scatter in the ages of elliptical galaxies is around
10\%.
The next obvious question to ask is whether metallicity variations amongst
these galaxies
would increase the scatter by a substantial amount.

To address this question, I have chosen to include a very simple prescription
for chemical
evolution in the models. The aim of the
exercise is to see whether the scatter in Mg$_{2}$ can still match observations
when metallicity
is allowed to vary according to the star formation history of a galaxy.
The assumptions of the model are as follows:
All stars that form produce a certain yield $y$ of heavy
elements, which is then added to the cold gas component of the galaxy. It is
assumed that
all metals are mixed instantaneously and uniformly with the cold gas and that
new stars that
form will have the metallicity of this enriched gas. The injection of cold gas
back into
the surrounding halo by supernova feedback will in turn enrich the hot
intergalactic medium.
In practice, supernovae explosions may be energetic enough to inject a certain
fraction of heavy elements
directly into the halo, without any mixing into the cold gas taking place.
Metal enrichment of the hot halo
gas is ignored
here as it would introduce  additional parameters into the model.
The yield $y$ is considered a free parameter and is adjusted so that the
central galaxy of a halo
with circular velocity $V_c= 220$ km s$^{-1}$ will have solar metallicity on
average.

Figures 5a and b show the distribution in metallicity and Mg$_{2}$ index for
elliptical galaxies
in both clusters and groups. There is a clear trend for fainter galaxies to
have lower
metallicities, but it is not strong enough to account for the strong
correlation
between Mg$_2$ and velocity dispersion seen in the data. The solid line in
figure 5b is the
Mg$_{2}$-$\sigma$ relation derived by BBF, where $\sigma$ has been converted
into V-magnitude
using the Faber-Jackson relation : $ (L/L_*) = (\sigma/ 220$ km s$^{-1})^4$ and
$M_*(V)=-20.7$.
One obvious way to steepen the correlation between Mg$_{2}$ and $\sigma$ would
be to assume that
more massive galaxies are more easily able to retain heavy elements
liberated by supernova explosions and reprocess them in stars. In less massive
galaxies, the energy of
the explosion is sufficient to cause much of this material
to be ejected into the intergalactic medium. Nonetheless it is encouraging that
the scatter obtained
in the Mg$_{2}$ relation is in good agreement with observations. At a fixed
V-band absolute
magnitude, galaxies brighter than $M_V=-20$
have a rms standard deviation $\simeq$ 0.02. For fainter systems, the rms
scatter increases to $\simeq$0.03.

\section {Conclusions}

In this letter, I have calculated the scatter in the mean ages of the stellar
populations of early-type galaxies, as predicted by a model in which
ellipticals and bulges
form by mergers of spiral disks in a universe where structure grows by a
process of
hierarchical clustering. The V-luminosity weighted mean stellar ages of
ellipticals in
the model lie in the range 8 to 12.5 Gyr, implying that the bulk of the stars
in these
galaxies were formed at redshifts exceeding 1.9. The scatter in the mean ages
of the model
E and SO galaxies is $\simeq$ 10-15 \%, well within the limits set by the data
of Bender, Burstein
\& Faber (1993) {\em provided} the intrinsic dispersion in the Mg$_{2}$ index
of observed
ellipticals is attributed solely to dispersion in age. The predicted scatter in
the U-V and
V-K colours of cluster ellipticals is less than the upper limit in the
intrinsic scatter of elliptical
galaxies in the Coma and Virgo clusters quoted by Bower, Lucy \& Ellis (1992).
The addition of later types increases the scatter substantially and again this
seems to agree with
observations.
Finally, I have investigated the effect of metallicity variation by
incorporating a very simple
prescription for chemical evolution into the model. The metallicity of galaxies
then varies
a result of differences in their star formation histories. Although this model
is not definitive,
it is encouraging that the mean
scatter in $Mg_{2}$ is still within observational bounds.
The main conclusion of this work, therefore, is that the observed
``homogeneity'' in the stellar
populations of elliptical galaxies does not pose any real problem for the
merger picture.
\\

\vspace{0.8cm}

\large
{\bf Acknowledgments}\
\normalsize
I thank Ralph Bender and Steve Zepf for helpful discussions.

\pagebreak

\vspace* {1cm}

\large
\begin{center} Table 1. Mean stellar age and rms standard deviation   \\
\vspace{0.6cm}
\normalsize
\begin {tabular}{llll}
{\bf environment and type} & {\bf Mean Age (Gyr)} & {\bf $\sigma$ rms} \\
cluster Es & 10.97 & 1.14 \\
cluster E+SOs & 10.55 & 1.46 \\
group Es & 10.61 & 1.11 \\
group E+SOs & 10.17 & 1.33 \\
field Es & 10.59 & 0.85 \\
field E+SOs & 9.84 & 1.33 \\
\end{tabular}
\end {center}

\pagebreak
\Large
\begin {center} {\bf References} \\
\end {center}
\vspace {1.5cm}
\normalsize
\parindent -7mm
\parskip 3mm

Barnes, J.E. \& Hernquist,L. 1992, Ann.Rev.Astr.Ap., 30, 705

Bender, R., Burstein, D. \& Faber, S.M. 1993, ApJ, 411, 153 (BBF)

Bond, J.R., Cole, S., Efstathiou, G. \& Kaiser, N. 1991, ApJ, 379, 440

Bower, R. 1991, MNRAS, 248, 332

Bower, R., Lucey, J.R. \& Ellis, R.S. 1992, MNRAS, 254, 601

Bruzual, G. \& Charlot, S. 1993, ApJ, 405, 538

Burstein, D., Davies, R.L., Dressler, A., Faber, S.M., Stone, R.P.S.,
Lynden-Bell, D.,
Terlevich, R.J. \& Wegner, G. 1987, ApJ, 313, 42

Dressler, A. 1984, ApJ, 281, 512

Dressler, A., Lynden-Bell, D., Burstein, D., Davies, R.L., Faber, S.M.,
Terlevich, R.J. \&
Wegner, G., 1987, ApJ, 313, 42

Faber, S.M. 1973, ApJ, 179, 423

Faber, S.M., Trager, S.C., Gonzalez, J.J. \& Worthey, G., 1995, preprint

Griffiths, R.E. {\em et al} 1995, ApJ, in press

Joseph, R.D. 1990, in Dynamica and Interactions of Galaxies, ed Wielen, R. (New
York:Springer), 132

Kauffmann, G. \& White, S.D.M. 1993, MNRAS, 261, 921

Kauffmann, G., White, S.D.M. \& Guiderdoni, B. 1993, MNRAS, 264, 201 (KWG)

Kauffmann, G. 1995, MNRAS, in press

Kennicutt, R.C. 1983, ApJ, 272, 54

Renzini, A. 1995, in Stellar Populations, ed Gilmore, G. \& van der Kruit, P.
(Dordrecht:Kluwer),
in press

Schweizer, F. \& Seitzer, P., 1992, AJ, 104, 1039

Simien, F. \& De Vaucouleurs, G. 1986, ApJ, 302, 564

Terlevich, R., Davies, R.L., Faber, S.M. \& Burstein, D. 1981, MNRAS, 196, 381

\pagebreak

\Large
\begin {center} {\bf Figure Captions} \\
\end {center}
\vspace {1.5cm}
\normalsize
\parindent 7mm
\parskip 8mm

{\bf Figure 1:} The V-luminosity weighted mean stellar ages of the galaxies in
two
clusters of mass $10^{15} M_{\odot}$. Large filled circles represent
elliptcals, open
squares are SOs, small filled circles are spirals and 3-pronged pinwheels are
elliptical
galaxies that were formed my a major merger in the past Gigayear.

{\bf Figure 2:} The mean ages of the early-type galaxies in a) halos with
masses of
$10^{12}-10^{13} M_{\odot}$. b) halos of mass $10^{14} M_{\odot}$.
Symbols are as in figure 1.

{\bf Figure 3:} The fraction of the total V-band light of cluster elliptical or
spiral galaxies contributed by stars which form
at a given epoch. The solid line is the average star formation history.
The squares and circles represent the maximum and minimum contributions
to the total light for any galaxy at that epoch.
The dotted line is a histogram of the percentage of elliptical galaxies that
experienced their
last major merger at a given epoch.

{\bf Figure 4:}The distribution in  U-V and V-K colour of early-type galaxies
in a
cluster of $10^{15} M_{\odot}$. Filled circles are ellipticals and open squares
are SOs.

{\bf Figure 5:} a) The metallicities of early-type galaxies in clusters derived
from
a simple chemical enrichment models (see text). b) The distribution in Mg$_{2}$
index.
Filled circles are ellipticals and open squares are SOs.

\end {document}